\documentclass[twocolumn,tighten]{aastex61}

\received{}
\revised{}
\accepted{}
\submitjournal{ApJ}

\shorttitle{$^3$H\lowercase{e}-Rich SEPs in Helical Jets}
\shortauthors{Bu\v{c}\'ik et al.}
\begin{document}

\title{$^3$H\lowercase{e}-Rich Solar Energetic Particles in Helical Jets on the Sun}

\correspondingauthor{Radoslav Bu\v{c}\'ik}
\email{bucik@mps.mpg.de}

%\author[0000-0002-0786-7307]{Greg J. Schwarz}
%\affil{American Astronomical Society \\
%2000 Florida Ave., NW, Suite 300 \\
%Washington, DC 20009-1231, USA}

%\author[0000-0001-7381-6949]{Radoslav Bu\v{c}\'ik}
\author{Radoslav Bu\v{c}\'ik}
\affiliation{Institut f\"{u}r Astrophysik, Georg-August-Universit\"{a}t G\"{o}ttingen, D-37077 G\"{o}ttingen, Germany}
\affiliation{Max-Planck-Institut f\"{u}r Sonnensystemforschung, D-37077 G\"{o}ttingen, Germany}

\author{Davina E. Innes}
\affiliation{Max-Planck-Institut f\"{u}r Sonnensystemforschung, D-37077 G\"{o}ttingen, Germany}
\affiliation{Max-Planck/Princeton Center for Plasma Physics, Princeton, NJ 08540, USA}

\author{Glenn M. Mason}
\affiliation{Applied Physics Laboratory, Johns Hopkins University, Laurel, MD 20723, USA}

\author{Mark E. Wiedenbeck}
\affiliation{Jet Propulsion Laboratory, California Institute of Technology, Pasadena, CA 91109, USA}

\author{Ra\'ul G\'omez-Herrero}
\affiliation{Space Research Group, University of Alcal\'a, E-28871 Alcal\'a de Henares, Spain}

\author{Nariaki V. Nitta}
\affiliation{Lockheed Martin Advanced Technology Center, Palo Alto, CA 94304, USA}



\begin{abstract}

Particle acceleration in stellar flares is ubiquitous in the Universe, however, our Sun is the only astrophysical object where energetic particles and their source flares can both be observed. The acceleration mechanism in solar flares, tremendously enhancing (up to a factor of ten thousand) rare elements like $^3$He and ultra-heavy nuclei, has been puzzling for almost 50 years. Here we present some of the most intense $^3$He- and Fe-rich solar energetic particle events ever reported. \added{The events were accompanied by non-relativistic electron events and type III radio bursts.} The corresponding high-resolution, extreme-ultraviolet imaging observations have revealed for the first time a helical structure in the source flare with a jet-like shape. \added{The helical jets originated in relatively small, compact active regions, located at the coronal hole boundary.} A mini-filament at the base of the jet appears to trigger these events. The events were observed with the two {\sl Solar Terrestrial Relations Observatories} STEREO on the backside of the Sun, during the period of increased solar activity in 2014. \deleted{During the last decade, it has been established that the helical motions in coronal jets represent propagating Alfv\'en waves. Revealing such magnetic-untwisting waves in the solar sources of highly enriched events in this study is consistent with a stochastic acceleration mechanism.}\added{The helical jets may be a distinct feature of these intense events that \replaced{produce}{is related to the production of} high $^3$He and Fe enrichments.}

\end{abstract}




\section{Introduction} \label{sec:intro}

Particle acceleration in solar flares \citep{mil98,lin11} remains an outstanding question in astrophysics. The elemental composition of the accelerated population differs remarkably from the abundances in the solar atmosphere \citep{mas07}. The composition is also distinct from the population accelerated by coronal mass ejection (CME) driven shocks in the solar atmosphere or interplanetary space \citep{rea13,rea15}. 

The most striking feature of particle acceleration in solar flares is the enormous enrichment of the rare isotope $^3$He by factors up to 10$^4$ above coronal or solar wind values \citep{koc84}. This $^3$He-rich solar energetic particle (SEP) population is further characterized by enrichments in heavy (Ne--Fe) and ultra-heavy (mass$>$70) ions by a factor of $\sim$2--10 and $>$100, respectively, independent of the amount of $^3$He enhancement \citep{mas86,mas04,rea94}. It has been interpreted as evidence that different mechanisms are involved in the acceleration of the $^3$He and the heavy nuclei \citep[however see][for possible exceptions]{mas16}. Whereas with He the lighter isotope is enhanced, with Ne and Mg the heavier isotopes, $^{22}$Ne and $^{26}$Mg, are enhanced by factors of $\sim$3 \citep{mas94,dwy01,wie10}. 

Flares associated with $^3$He-rich SEP events are commonly observed as collimated or jet-like form in extreme-ultraviolet (EUV) or X-ray images \citep{nit06,nit08,nit15,buc14,buc15,buc16,che15}. Sometimes jets show a high-altitude extension in white-light coronograph observations \citep{kah01,wan06}. The association with coronal jets has been interpreted as evidence for magnetic reconnection involving field lines open to interplanetary space \citep{shi00}. Reconnecting magnetic field impulsively releases energy which is available for particle acceleration \citep{ben10}. The underlying photospheric source of $^3$He-rich SEPs is typically a small, compact active region often found next to a coronal-hole \citep{wan06,buc14}. $^3$He-rich SEPs are almost always accompanied by type III radio bursts \citep{rea86,nit06}, the emission excited by energetic electrons streaming through the corona and into interplanetary space \citep{kru99}. 

Characteristic features of jets in $^3$He-rich SEP events \citep{inn16} have not been systematically investigated. Depending on their morphology, solar jets have been recently divided into two categories, standard and blowout (or erupting) jets \citep{rao16}. With their peculiar untwisting motion, blowout jets are still not well understood. It is thought that they involve more complex reconnection processes than standard, straight jets. More importantly, the unwinding motions in the jets have been associated with generation and propagation of Alfv\'en waves \citep{cir07,nis08,par09,tor09,liu11,sch13,moo15,lee15}. A resonant interaction with plasma turbulence (including Alfv\'en waves) has been often involved in models of $^3$He-rich SEPs \citep{fis78,tem92,mil98,zha04,liu06,eic14,kum17}. In addition, recent theoretical work suggests that helicity in the magnetic field, though not directly applied on jets, may play a critical role in particle acceleration \citep{fle13,gor14}. In this paper, we examine remarkably intense $^3$He-rich SEP events and find their solar source associated with erupting EUV jets clearly showing helical motion. This finding may have a broader application - it is the first observation of rotating magnetized plasma flows in a stellar flare being associated with energetic ion acceleration.

\section{Methods} \label{sec:met}

The $^3$He-rich SEP events reported in this paper were identified using observations from the time-of-flight mass spectrometer Suprathermal Ion Telescope (SIT) \citep{mas08} on the twin {\sl Solar Terrestrial Relations Observatories} AHEAD (STEREO-A) and BEHIND (STEREO-B) spacecraft. The SIT has a sunward viewing direction close to the average Parker magnetic field spiral line. The instrument measures element composition from He to Fe in the kinetic energy range from 20\,keV\,nucleon$^{-1}$ to several MeV\,nucleon$^{-1}$. We also make use of energetic electron measurements made by the STE-D \citep{lin08} and the SEPT \citep{mul08} sensors, solar wind measurements made by the PLASTIC instrument \citep{gal08}, and interplanetary magnetic field measurements obtained by the magnetometer \citep{acu08} on STEREO. Notice that data from the sunward pointing STE-U are not available because of instrument saturation by sunlight \citep{wan12}. 

The solar sources of $^3$He-rich SEPs were examined using high-resolution full-disk STEREO EUV images from the SECCHI/EUVI instrument \citep{how08}. The EUVI instrument observes the Sun in four wavelengths (304, 171, 195, 284\,{\AA}), and in 2048$\times$2048 1.6 arc second pixels in a circular field of view which extends to 1.7\,$R_{\sun}$. The 304, 171, 195 and 284\,{\AA} channels observe emissions from He~II (temperature T$\sim$0.05\,MK), Fe~IX (T$\sim$0.7\,MK), Fe~XII (T$\sim$1.6\,MK) and Fe~XV (T$\sim$2.2\,MK) lines, respectively. They provide snapshots of chromospheric (304\,{\AA}), the upper transition region (171\,{\AA}) and coronal plasma (195, 284\,{\AA}). The channels 304 and 171\,{\AA} had a high temporal resolution for the investigated events, 150\,s and 75\,s, respectively. Occasionally, 171\,{\AA} images were obtained with even higher resolution (30--45\,s). We also inspect dynamic radio spectra for the event associated type III radio bursts. The radio data are provided by the WAVES/STEREO instrument \citep{bou08} with a frequency range ($<$16\,MHz) covering emission generated from about 2\,$R_{\sun}$ to 1\,AU. 

\begin{figure}
%\figurenum{1}
%\epsscale{0.7}
\plotone{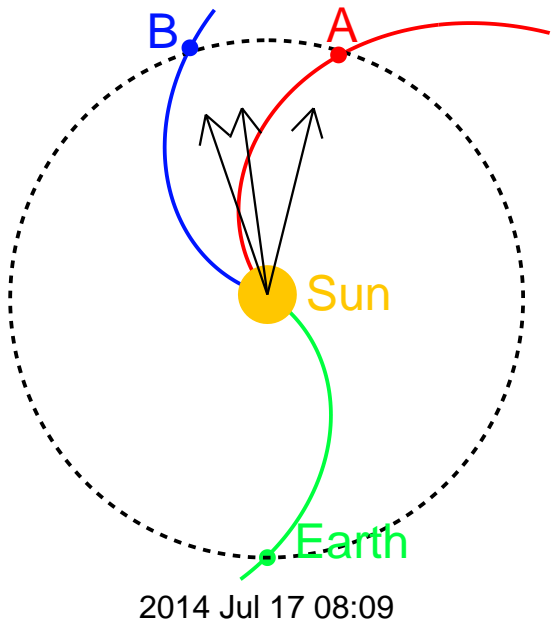}
\caption{Ecliptic plane projection of the STEREO-B, STEREO-A, and the Earth locations on 2014 July 17 08:09\,UT. Parker spirals for 400\,km\,s$^{-1}$ solar wind speed are shown. The arrows indicate the longitude of the source active regions on the Sun in event 2 on 2014 July 17 08:09\,UT (the rightmost arrow), event 3 on July 19 19:30\,UT (the leftmost arrow) and event 1 on April 29 19:30\,UT (the middle arrow).\deleted{The nominal magnetic connection of STEREO-A was closer to the source active regions than the connection of STEREO-B.} Note that the angular separation between the two STEREOs differs only by 4$^{\circ}$ on April 29 and therefore the overall constellation remains similar. \label{fig:f1}}
\end{figure}

The location of the two STEREOs during the investigated events is schematically shown in Figure \ref{fig:f1}. Both STEREOs were in a heliocentric orbit at $\sim$1\,AU near the ecliptic plane separated from the Earth by $\sim$160$^{\circ}$ (STEREO-B between $-$162$^{\circ}$ and $-$165$^{\circ}$, and STEREO-A between 157$^{\circ}$ and 164$^{\circ}$ of heliographic longitude). The angular separation between the two STEREOs was between 34$^{\circ}$ and 38$^{\circ}$. \added{The arrows in Figure \ref{fig:f1} indicate the longitude of the source active regions on the Sun in the examined events. The nominal magnetic connection of STEREO-A was closer to the source active regions than the connection of STEREO-B.}

\section{Results} \label{sec:res}
\subsection{$^3$He-rich SEP events} \label{subsec:eve}

\begin{figure*}
%\figurenum{2}
%\epsscale{1.1}
\plotone{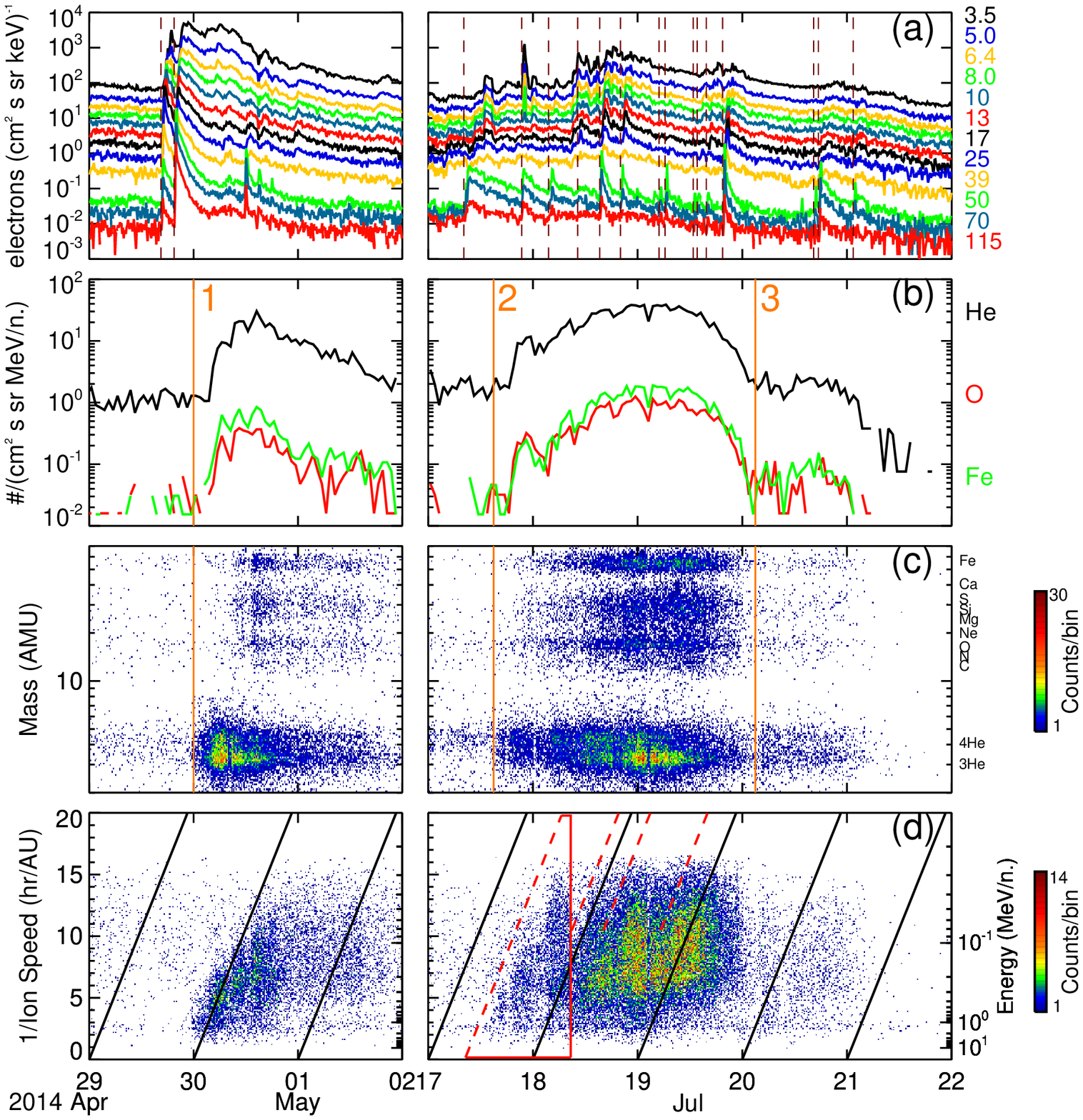}
\caption{(a) 10-min electron intensities from STE-D/STEREO-A (3.5--39\,keV) anti-sunward and SEPT/STEREO-A (50--115\,keV) sunward pointing sensors. Dashed vertical lines indicate EUVI flares listed in Table \ref{tab:tab2}. (b) 1 hr SIT/STEREO-A 160--226\,keV\,nucleon$^{-1}$ He\deleted{(3--8\,amu)}, O\deleted{(13--19\,amu)}, Fe\deleted{(36--80\,amu)} intensities. (c) \added{80--900\,keV\,nucleon$^{-1}$}SIT/STEREO-A\deleted{binned} mass spectrogram\deleted{of individual pulse-height analyzed ions in the energy ranges 250--900\,keV\,nucleon$^{-1}$ (mass$<$8\,amu) and 80--150\,keV\,nucleon$^{-1}$ (mass$>$8\,amu)}.\deleted{The color code indicates the number of ions per bin, where low values correspond to blue and high to red color.} Solid vertical lines in panels (b--c) mark start times of the ion events\deleted{at STEREO-A}. (d) SIT/STEREO-A spectrogram of 1/(ion-speed) versus arrival times of 10--70\,amu ions. Sloped solid lines indicate arrival times for ions\deleted{leaving the Sun at 0\,UT and} traveling along a spiral of 1.13\,AU length, without scattering. \deleted{The field line length is calculated for STEREO-A at heliocentric distance 0.96\,AU and measured solar wind speed median 350\,km\,s$^{-1}$.}The slanted dashed lines approximately mark four ion injections with the velocity dispersion in event 2 (trapezoid marks the integration limits for the first injection). \label{fig:f2}}
\end{figure*}

Figure \ref{fig:f2} shows three examined $^3$He-rich SEP events observed on STEREO-A on 2014 April 30 (number 1), July 17 (number 2) and July 20 (number 3), marked by vertical lines in panels (b--c). Note that the July events (2, 3) were associated with the same active region on the Sun. \added{Figure \ref{fig:f2}a presents electron intensities at different energy channels between 3.5 and 115\,keV. Figure \ref{fig:f2}b shows He (3--8\,amu), O (13--19\,amu), and Fe (36--80\,amu) intensities at 160--226\,keV\,nucleon$^{-1}$.} The \added{$^3$He-rich}events were accompanied by multiple solar energetic electron events\deleted{(see Figure \ref{fig:f2}a) with associated type III radio bursts}. A remarkable number of electron injections (at least fifteen) were observed within 3.4 days in the July $^3$He-rich SEP events 2 and 3. The corresponding multiple ion injections are not well resolved in the time-intensity profiles in Figure \ref{fig:f2}b. Instead, the compound profiles show a gradual rise during $\sim$1.5 days and a relatively rapid decay within $\sim$0.5 days in event 2. \added{Figure \ref{fig:f2}c presents mass spectrogram of all individual pulse-height analyzed ions in the energy ranges 250--900\,keV\,nucleon$^{-1}$ (mass$<$8\,amu) and 80--150\,keV\,nucleon$^{-1}$ (mass$>$8\,amu).} The mass spectrogram \deleted{in Figure \ref{fig:f2}c}shows a clear dominance of the $^3$He isotope. Event 1 \replaced{shows}{has} a clear velocity dispersive onset\added{, where higher energy ions arrived earlier than lower energy ones, as shown by the triangular pattern in the inverted ion-speed versus time spectrogram in Figure \ref{fig:f2}d.} \added{Sloped solid lines in Figure \ref{fig:f2}d indicate \replaced{arrival times}{velocity dispersion} for ions leaving the Sun at 00:00\,UT and traveling along a spiral of 1.13\,AU length, without scattering. The field line length is calculated for STEREO-A located at heliocentric distance 0.96\,AU and measured solar wind speed median 350\,km\,s$^{-1}$.} For event 2 four overlapping ion injections with velocity dispersion are discernable\deleted{in the inverted ion-speed spectrogram} in Figure \ref{fig:f2}d: middle and end of July 17, beginning and middle of July 18. These four ion injections are marked by slanted dashed lines that approximately follow the inclined pattern in the ion spectrogram. \added{The first and to a lesser degree also the third ion injection show an intermittent time profile, containing short periods with decreased ion counts. This has been attributed to the encounters of field lines not populated with ions from on-going injection on the Sun \citep[e.g.,][]{maz00}.} No velocity \replaced{dispersion}{dispersive onset} is seen in \added{Figure \ref{fig:f2}d for} event 3\added{; an increase of the registered counts occurs at all energies at the same time}.

\begin{figure}
%\figurenum{3}
%\epsscale{0.7}
\plotone{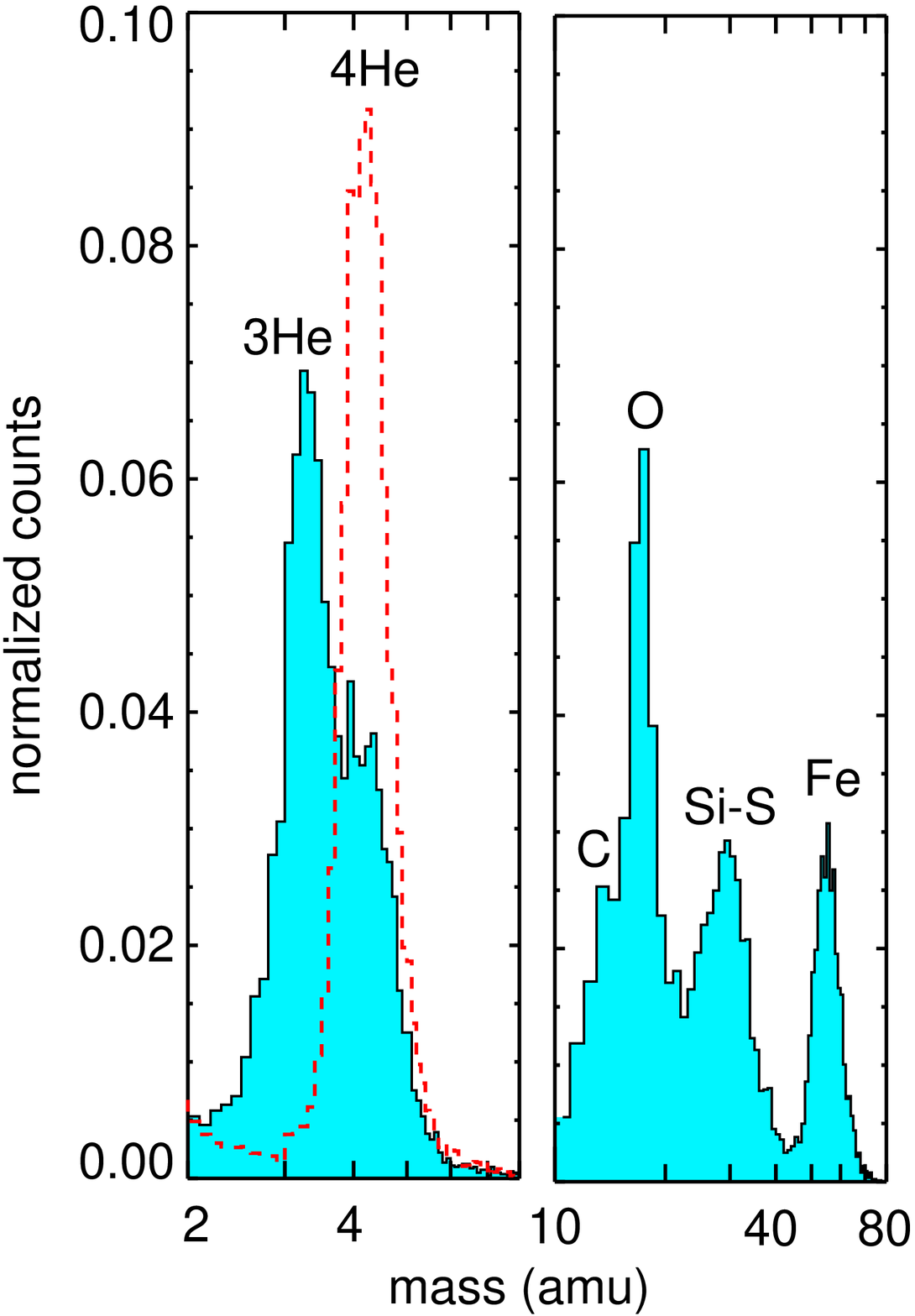}
\caption{Mass histograms (normalized to the total number of counts) at 320--450\,keV\,nucleon$^{-1}$ (2--8\,amu) and 80--113\,keV\,nucleon$^{-1}$ (10--80\,amu) for event 2 (July 17). Red histogram is from SIT observations of a clean $^4$He peak\deleted{, used to evaluate $^3$He/$^4$He ratio. The number of counts under the Fe mass peak is higher than under O peak, though it may not be obvious on the logarithmic mass scale. Folding in the detection efficiencies of O (60\%) and Fe (89\%) at 80--113\,keV\,nucleon$^{-1}$, the Fe/O ratio results in $\sim$0.9}. \label{fig:f3}}
\end{figure}

Figure \ref{fig:f3} shows SIT mass histograms for helium \added{(at 320--450\,keV\,nucleon$^{-1}$)} and heavier ions \added{(at 80--113\,keV\,nucleon$^{-1}$)} in event 2 (July 17) used for measuring abundances. \added{Over-plotted is a histogram from SIT observations of a clean $^4$He peak, used to evaluate $^3$He/$^4$He ratio. Note that the number of counts under the Fe mass peak is higher than under O peak, though it may not be obvious on the logarithmic mass scale. Folding in the detection efficiencies of O (60\%) and Fe (89\%) at 80--113\,keV\,nucleon$^{-1}$, the Fe/O ratio results in $\sim$0.9.} Key elemental ratios at 320--450\,keV\,nucleon$^{-1}$, the energy bin where the abundances of suprathermal ions have been quoted, are listed in Table \ref{tab:tab1}. The table further shows the start and end times of the $^3$He-rich periods, and the $^3$He and Fe fluences over the spectral ranges $\sim$0.2--2.0 and $\sim$0.1--1.0\,MeV\,nucleon$^{-1}$, respectively. The values for two, the least overlapped ion injections in event 2 are also listed.

\begin{deluxetable*}{clcccccc}
\tablecaption{STEREO-A $^3$He-rich SEP events: Energetic ion properties \label{tab:tab1}}
\tablehead{
\colhead{Event} & \colhead{Start date} & \colhead{Days} & \colhead{$^3$He/$^4$He\tablenotemark{a}} & \colhead{Fe/O\tablenotemark{a}}  & \colhead{$^3$He fluence\tablenotemark{b}} & \colhead{Fe fluence\tablenotemark{b}} & \colhead{Notes}\\
\colhead{Number} & \colhead{} & \colhead{} & \colhead{} & \colhead{} & \colhead{($\times$10$^{3}$)} & \colhead{($\times$10$^{3}$)} & \colhead{}
}
%\colnumbers
\decimals
\startdata
1 & 2014 Apr 30 & 120.0--122.0 & 1.75$\pm$0.07 & 2.18$\pm$0.17 & \phn76.4$\pm$1.1\phn   & 13.7\phn$\pm$0.2\phn & \nodata \\
2 & 2014 Jul 17  & 198.6--201.0 & 1.46$\pm$0.05 & 1.58$\pm$0.08 & 112$\pm$1\phn       & 45.5\phn$\pm$0.5\phn & \nodata \\
   & 2014 Jul 17  & 198.3--199.4 & 0.35$\pm$0.04 & 1.45$\pm$0.18 & \phn9.16$\pm$0.46 & \phn4.25$\pm$0.14 & Inject. 1 \\
   & 2014 Jul 19  & 200.1--201.0 & 3.76$\pm$0.27 & 1.45$\pm$0.12 & \phn53.7$\pm$0.8\phn & 20.6\phn$\pm$0.3\phn & Inject. 4 \\
3 & 2014 Jul 20  & 201.1--202.1 & 1.15$\pm$0.16 & 1.82$\pm$0.43 & \phn7.14$\pm$0.36 & \phn1.68$\pm$0.09 & \nodata \\
\enddata
\tablenotetext{a}{0.320--0.450\,MeV\,nucleon$^{-1}$\added{; $^3$He/$^4$He$\sim$0.15 and Fe/O$\sim$1.13 are typical values at this energy (see text for details)}}
\tablenotetext{b}{0.226--1.810\,MeV\,nucleon$^{-1}$ for $^3$He; 0.113--0.905\,MeV\,nucleon$^{-1}$ for Fe; fluence unit -- particles (cm$^2$\,sr\,MeV/nucleon)$^{-1}$}
\end{deluxetable*}

\begin{figure*}
%\figurenum{4}
%\epsscale{1.}
\plotone{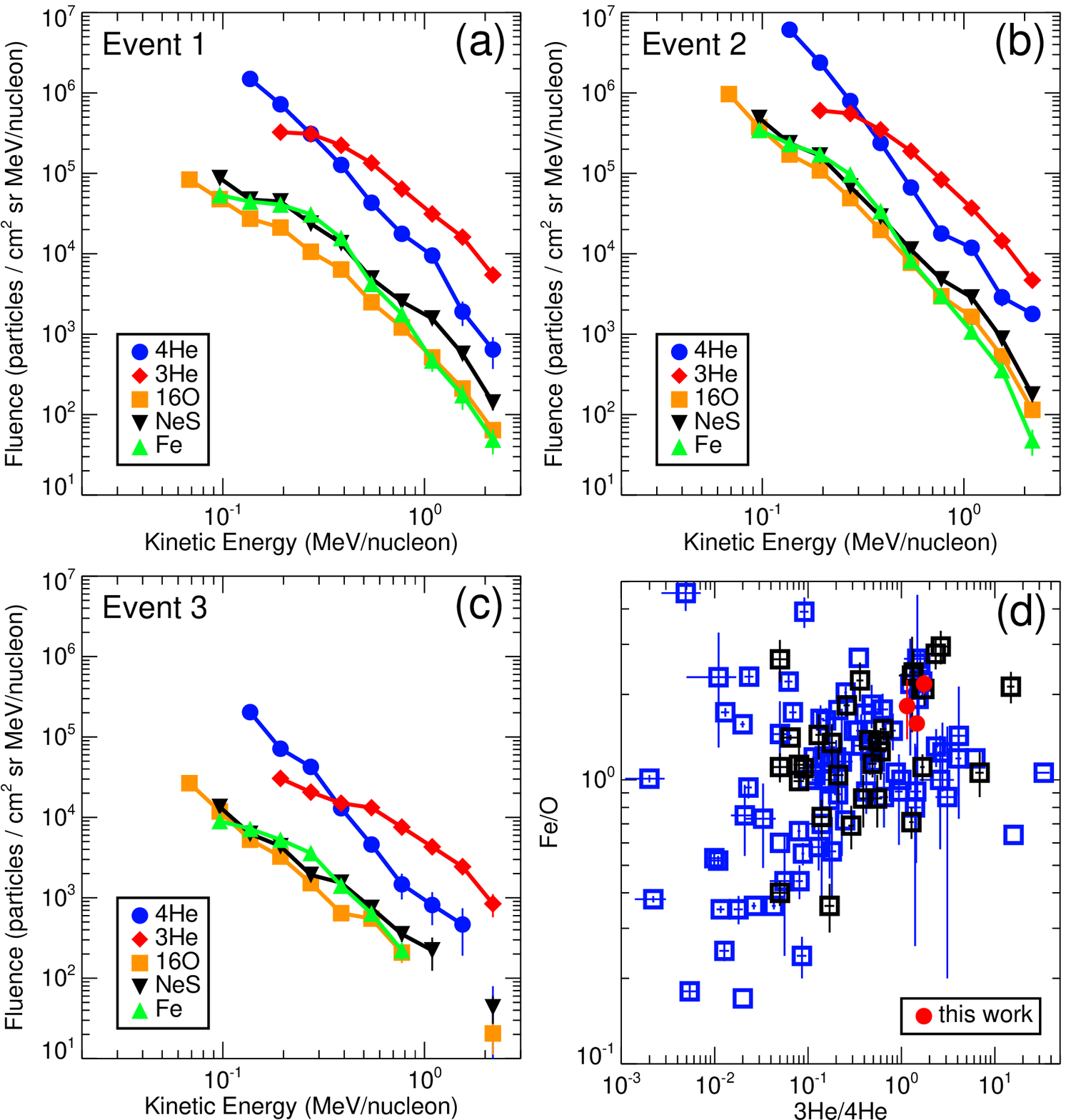}
\caption{(a--c) Differential energy spectra for event 1 (April 30), 2 (July 17), 3 (July 20)\deleted{for $^3$He, $^4$He, O, NeS, and Fe} as observed by SIT on STEREO-A.\deleted{The individual mass peaks in the range Ne--S ($^{20}$Ne, $^{24}$Mg, $^{28}$Si, $^{32}$S) are not resolved with SIT.} (d) Fe/O versus $^3$He/$^4$He for 109 previously reported $^3$He-rich SEP events. Blue squares mark the $^3$He/$^4$He at 320--450\,keV\,nucleon$^{-1}$\deleted{\citep{mas02,mas04,wan06,buc13,buc14,buc16}} or 400--600\,keV\,nucleon$^{-1}$\deleted{\citep{wan06,pic06}}. Black squares mark the $^3$He/$^4$He at 0.5--2\,MeV\,nucleon$^{-1}$\deleted{\citep{tyl02,nit15,mas16}}. The Fe/O is predominantly\deleted{(91 cases)} from 320--450\,keV\,nucleon$^{-1}$; the remaining Fe/O is from 400--600 or 160--226\,keV\,nucleon$^{-1}$. Red circles mark events 1, 2 and 3 in this study. \label{fig:f4}}
\end{figure*}

Figure \ref{fig:f4}(a--c) shows the event-integrated \added{$^3$He, $^4$He, O, NeS, and Fe} fluence spectra for the range $\sim$0.07--2\,MeV\,nucleon$^{-1}$ with the background and the helium spillover subtracted. \added{The individual mass peaks in the range Ne--S ($^{20}$Ne, $^{24}$Mg, $^{28}$Si, $^{32}$S) are not resolved with SIT.} The approximate event periods were identified using the 250--900\,keV\,nucleon$^{-1}$ He measurements in the mass spectrograms in Figure \ref{fig:f2}c. The fractions of ions that spill into the $^3$He or $^4$He peaks were estimated on the basis of the SIT observations of a clean $^4$He peak \added{(see Figure \ref{fig:f3})} and the cleanest $^3$He peak \citep{buc13}. The background with a signature at $\sim$1\,MeV\,nucleon$^{-1}$ is due to cosmic rays penetrating the SIT telescope housing. After correction for the background, some high-energy spectral points in event 3 were omitted due to low statistical accuracy. A common feature of the observed spectra is a distinct shape for $^3$He and Fe as it has been previously reported in some $^3$He-rich SEP events  \citep[class-2 events in][]{mas00}. The $^3$He spectra appear to have a spectral break at $\sim$500\,keV\,nucleon$^{-1}$ and the Fe spectra at $\sim$300\,keV\,nucleon$^{-1}$. For events 1 and 2, it is also seen in NeS. Figure \ref{fig:f4}d shows Fe/O versus $^3$He/$^4$He obtained from all (109) previously reported $^3$He-rich SEP events at suprathermal ($\lesssim$1\,MeV\,nucleon$^{-1}$) energies\deleted{\citep{tyl02,mas02,mas04,mas16,wan06,pic06,buc13,buc14,buc16,nit15}}. \added{Specifically, the $^3$He/$^4$He ratio is reported at 320--450\,keV\,nucleon$^{-1}$ in 67 events \citep{mas02,mas04,wan06,buc13,buc14,buc16}, at 400--600\,keV\,nucleon$^{-1}$ in 11 events \citep{wan06,pic06}, and at 0.5--2\,MeV\,nucleon$^{-1}$ in 31 events \citep{tyl02,nit15,mas16}. The Fe/O is predominantly (91 events) given at 320--450\,keV\,nucleon$^{-1}$; the remaining Fe/O is given at 400--600 or 160--226\,keV\,nucleon$^{-1}$. The corresponding $^3$He/$^4$He and Fe/O median values at 320--450\,keV\,nucleon$^{-1}$ are 0.15 and 1.13, respectively.} Over-plotted \added{in Figure \ref{fig:f4}d} are the \added{$^3$He/$^4$He and Fe/O} ratios for events 1--3 in this study. The examined events are strongly enhanced both in $^3$He and heavier elements that have not been often reported in the suprathermal energy range. Only a few from so far reported events have such large enrichment in both $^3$He and Fe: up to 5 events at 320--450\,keV\,nucleon$^{-1}$ and up to 6 events at 0.5--2\,MeV\,nucleon$^{-1}$. 

\begin{deluxetable*}{clclccccc}
\tablecaption{STEREO-A $^3$He-rich SEP events: Solar source properties \label{tab:tab2}}
\tablehead{
\colhead{Event} & \multicolumn{3}{c}{EUVI Flare} & \colhead{Type III} & \colhead{Electron} &\multicolumn{2}{c}{CME\tablenotemark{d}} & \colhead{Footpoint}\\
\cline{2-4}
\cline{7-8}
\colhead{Number} & \colhead{Start} & \colhead{Location\tablenotemark{a}} & \colhead{Type\tablenotemark{b}} & \colhead{start} & \colhead{event start\tablenotemark{c}} & \colhead{Speed} & \colhead{Width} & \colhead{Longitude\tablenotemark{a}}%\\
}
\startdata
1 & Apr 29 16:30 & W30S09 & B  & 16:30 & 16:52$\pm$1 min & 199 & 80 & W64 \\
   & Apr 29 19:30 & W31S09 & J+ & 19:30 & 19:50$\pm$1 min & 422 & 43 & W72 \\
2 & Jul 17  08:09 & W03S10 & J+ & 08:08 & 08:42$\pm$1 min & 332 & 120 & W68 \\
   & Jul 17  21:25 & W10S09 & B+ & 21:17 & 21:42$\pm$1 min & 479 & 28 & W68 \\
   & & & & 21:24 & & & & \\
   & Jul 18 03:35 & W13S10 & B+ & 03:40 & 04:18$\pm$5 min & 359 & 29 & W67 \\
%   & & & & multiple & & & & \\
   & Jul 18 10:15 & W17S10 & B   & 10:15 & 10:33$\pm$1 min & \nodata & \nodata & W57 \\
   & Jul 18 15:20 & W20S10 & J+  & 15:20 & 15:35$\pm$1 min  & 462 & 28 & W62 \\
   & & & & & 15:42$\pm$1 min & & & \\
   & Jul 18 20:05 & W22S10 & B   & 20:00 & 20:28$\pm$5 min & \nodata & \nodata & W56 \\
   & Jul 19 04:55 & W27S10 & B   & 04:50 & 05:08$\pm$5 min & \nodata & \nodata & W55 \\
   & Jul 19 06:18 & W28S10 & B   & 06:18 & 06:38$\pm$1 min & \nodata & \nodata & W57 \\
   & Jul 19 12:45 & W32S10 & B   & 12:44 & 13:08$\pm$5 min & \nodata & \nodata & W53 \\
   & Jul 19 13:40 & W32S10 & B   & 13:35 & 13:54$\pm$1 min & 396 & 10 & W54 \\
   & Jul 19 15:45 & W33S10 & B   & 15:45 & 16:08$\pm$5 min & 539 & 10 & W53 \\
3 & Jul 19 19:30 & W35S10 & J+  & 19:30 & 19:54$\pm$1 min & 509 & 32 & W49 \\   
   & Jul 20 16:21 & W47S11 & B    & 16:20 & 16:48$\pm$5 min & \nodata & \nodata & W64 \\ 
   & Jul 20 17:27 & W47S11 & J    & 17:27 & 17:54$\pm$1 min & \nodata & \nodata & W63 \\
%   & & & & multiple & & & & \\
   & Jul 21 01:25 & W52S10 & J   & 01:23 & 01:53$\pm$5 min & 568 & 19 & W57 \\
\enddata
\tablenotetext{a}{from STEREO-A point of view}
\tablenotetext{b}{B: brightening, J: jet; marked with + are flares associated with the $^3$He-rich SEP \replaced{events}{injection}}
\tablenotetext{c}{55--65\,keV SEPT/STEREO-A sunward pointing telescope; Jul 18 10:33\,UT from the south telescope}
\tablenotetext{d}{speed (km\,s$^{-1}$) and width ($^{\circ}$) from SOHO/LASCO catalog (\url{http://cdaw.gsfc.nasa.gov/CME_list}) }
\end{deluxetable*}

Two investigated events (number 1 and 2) show exceptionally high ion fluences. \citet{ho05} reported an upper limit of $^3$He fluence (2.0$\times$10$^5$\,particles (cm$^2$\,sr\,MeV\,nucleon$^{-1}$)$^{-1}$ at 0.2--2\,MeV\,nucleon$^{-1}$) examining 201 SEP events during a 6-year period around the solar maximum of the previous solar cycle 23. The authors found 97\% events with $^3$He fluence below 1.1$\times$10$^5$ and 94\% below 0.8$\times$10$^5$\,particles (cm$^2$\,sr\,MeV\,nucleon$^{-1}$)$^{-1}$ which are the $^3$He fluence values in event 2 and event 1, respectively (see Table \ref{tab:tab1}). The Fe fluence in the largest $^3$He-rich SEP event of the previous solar cycle (2001 April 14) \citep{tyl02} was a factor of $\sim$3--5 lower than the Fe fluences in the first five spectral points (97--386\,keV\,nucleon$^{-1}$) of event 2 and comparable to the fluences in event 1. The Fe/O \added{ratio} of 1.4 \added{in 2001 April 14 event} was similar to values in our events, but the $^3$He/$^4$He ratio attained only $\sim$6.5\% (c.f. 175\% in event 1) at 0.5--2\,MeV\,nucleon$^{-1}$. The high fluences in event 2 were caused by multiple (at least four) ion injections occurred within the span of 1.3 days. The first two ion injections were apparently weaker than the two later one. For instance, Table \ref{tab:tab1} indicates that the $^3$He fluence in the first and the fourth ion injections present $\sim$8\% and 48\% of the total $^3$He fluence of event 2, respectively.   

\subsection{Event sources} \label{subsec:sou}

Table \ref{tab:tab2} presents the solar source properties of the STEREO-A $^3$He-rich event periods described in this study. Column 1 gives the event number. Column 2 indicates the start time of the STEREO-A EUVI flare. Column 3 gives the flare location from the STEREO-A point of view, column 4 the EUV flare type (brightening, jet). Note, that four flares with EUV brightening were associated with jet-like CMEs (width 10$^{\circ}$--30$^{\circ}$) or white-light jets. As seen from the Earth the flares were between W166 and E145 in the July events and at E173 in the April event. Column 5 gives the type III radio burst start time. Column 6 indicates the SEPT/STEREO-A electron event onset times at energy 55--65\,keV (or speed $\sim$0.45c). Columns 7 and 8 indicate the CME speed and width, respectively, obtained from the SOHO LASCO CME Catalog. Column 9 gives the foot-point longitude of the Parker spiral connecting STEREO-A, determined from the measured solar wind speed at the type III radio burst onset time. 

Event 1, \replaced{with}{showing} a dispersive onset, was preceded by two type III radio bursts observed only 3 hours apart. Extrapolating backward in time, along the triangular pattern in the inverse velocity-time spectrogram in Figure \ref{fig:f2}d, favors the second type III burst and the associated jet on April 29 19:30\,UT as the most likely event source. With this technique, the $<$1\,MeV\,nucleon$^{-1}$ ion injection times can be estimated with an uncertainty of $\pm$45\,minutes \citep{mas00}. The ions in the energy range 160--226\,keV\,nucleon$^{-1}$ (displayed in Figure \ref{fig:f2}b) would arrive at STEREO-A within $\sim$7.1--8.5\,hours without scattering and with a zero-pitch angle. If injected with the type III burst observed on April 29 19:30\,UT, the expected earliest arrival time of ions from this energy range would be on April 30 $\sim$02:30\,UT which is consistent with the event onset at 03:30\,UT observed in the ion intensity-time profiles (see Figure \ref{fig:f2}b). Note that small electron events on April 30 (with no obvious ion injections) were associated with another active region. 

The velocity dispersion technique suggests that four ion injections in event 2, marked by the slanted dashed lines in Figure \ref{fig:f2}d, are associated with the EUV flares on July 17 08:09\added{\,UT} (jet), \added{July 17} 21:25\,UT (brightening),\deleted{and} July 18 03:35\added{\,UT} (brightening) \replaced{,}{and July 18} 15:20\,UT (jet). The association with a corresponding type III radio burst was quite straightforward for the first three ion injections as there were only three, well-separated in time radio bursts in the interval from July 17 00\,UT to July 18 09\,UT. The EUV flare type on July 17 21:25\,UT and July 18 03:35\,UT, \deleted{are}marked as brightening, \deleted{although these} were accompanied by white-light jets as observed in the LASCO coronograph (see Table \ref{tab:tab2}). \added{Thus, all four ion injections in event 2 were associated with jets. Note, there were other seven brightenings in the same active region during the event 2 \replaced{with no obvious ion injections}{without a clear jet shape. These brightenings were accompanied by energetic electrons and type III radio bursts, but no obvious ion injections.} \replaced{These brightenings are for}{For} completeness\added{, these events are} indicated in Table \ref{tab:tab2}.} Event 2 was also measured on STEREO-B (not shown), separated by 34$^{\circ}$ west from STEREO-A, but it was not well seen due to enhanced background from the prior events combined with the instrument mass resolution limitations. The first three solar electron injections in event 2 (July 17) were clearly observed with the electron telescopes on STEREO-B. 

\begin{figure*}
%\figurenum{5}
%\epsscale{1.}
\plotone{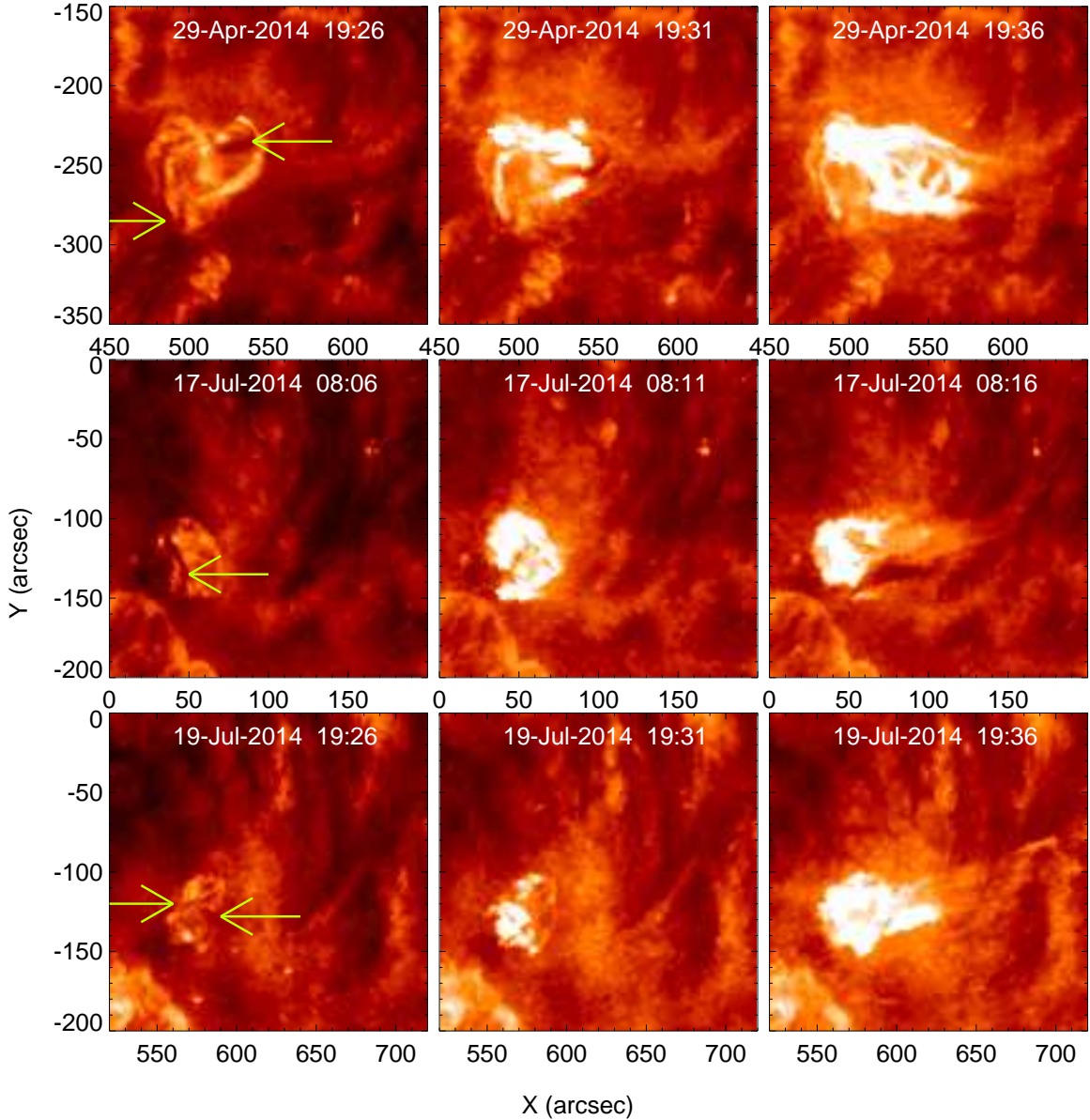}
\caption{STEREO-A EUV (304\,{\AA}) images of the solar source for event 1 (upper row), event 2 - the first ion injection (middle row) and event 3 (lower row).  (Left) pre-flare, (middle) flare onset, (right column) jet time. The arrows mark cool mini filaments. \label{fig:f5}}
\end{figure*}

The non-dispersive event 3 is most probably associated with the jet and type III radio burst on July 19 19:30\,UT, accompanied by an electron event on July 19 19:54\,UT. The electron event is dispersive and highly anisotropic with the highest intensities observed in the sunward pointing telescope. The expected earliest arrival time of ions from the 160--226\,keV\,nucleon$^{-1}$ energy range would be on July 20 $\sim$02:30\,UT. The He and Fe intensities in event 3 show an increase on July 20 $\sim$03:30\,UT (see Figure \ref{fig:f2}b) consistent with the ion injection at 19:30\,UT on the previous day. The EUV jet on July 20 17:27\,UT, associated with a quite intense solar electron event, was not accompanied by an energetic ion event that would have been observed on July 21. The lack of the ion detection is likely related to the disconnection with the solar source due to an encounter of a stream interaction region\footnote{An enhanced solar wind proton density and a gradual rise in solar wind speed were observed on July 21. This time period is included in the list of stream interaction regions on STEREO (\url{http://www-ssc.igpp.ucla.edu/forms/stereo/stereo_level_3.html}).}, though other effects like weak (or no) ion production at the source could not be ruled out. 

Figure \ref{fig:f5} shows STEREO-A EUV (304\,{\AA}) images of the solar sources for the $^3$He-rich SEP events under study. The images are shown at the pre-flare, onset and jet times in a 5-minute sequence. Note that flares in the middle (July 17, event 2) and lower (July 19, event 3) rows of Figure \ref{fig:f5} originate in the same active region. For event 2 we only show the source flare for the first ion injection (the July 17 08:09\,UT jet) where the high-cadence data are available. The source regions contained a small filament (dark bent strip) at their bases marked by arrows in Figure \ref{fig:f5}. The filament in event 1 shows vertical and horizontal components and its eruption at 19:31\,UT was clearly captured. The filament in event 2 is less extended, forming a wave-like shape in the vertical direction. A similar mini-filament is observed in 10-minute 304\,{\AA} images in the July 18 15:20\,UT jet\added{, associated with the fourth ion injection} of event 2. Overall, the configuration in the source for event 1 looks more complex with many intertwined loops. The filament in event 3 appears to be faint. A small filament of cool (T $\sim$0.01--0.10\,MK) plasma has been often observed in blowout jets \citep{moo10,hon11,she12,fil15}. 

%The source active region in event 1 emerged at the border of a near-equatorial coronal hole on April 26 $\sim$05\,UT as seen in EUV 304\,{\AA} images. Note %that photospheric magnetic field observations, where emerging active regions are usually well observed, are not available from the backside of the Sun. The %source region in events 2 and 3 is also located on the equatorial coronal hole border and emerged on July 16 around 12\,UT. The coronal holes, the regions with %magnetic field open to the heliosphere, were most clearly seen in 284\,{\AA} EUV images (see Figures \ref{fig:f7}a and \ref{fig:f7}b). Figures \ref{fig:f7}a and \ref{fig:f7}b demonstrate that the solar sources were relatively (compared to other bright areas on the disk) small and compact regions. 

\begin{figure*}
%\figurenum{6}
%\epsscale{1.}
\plotone{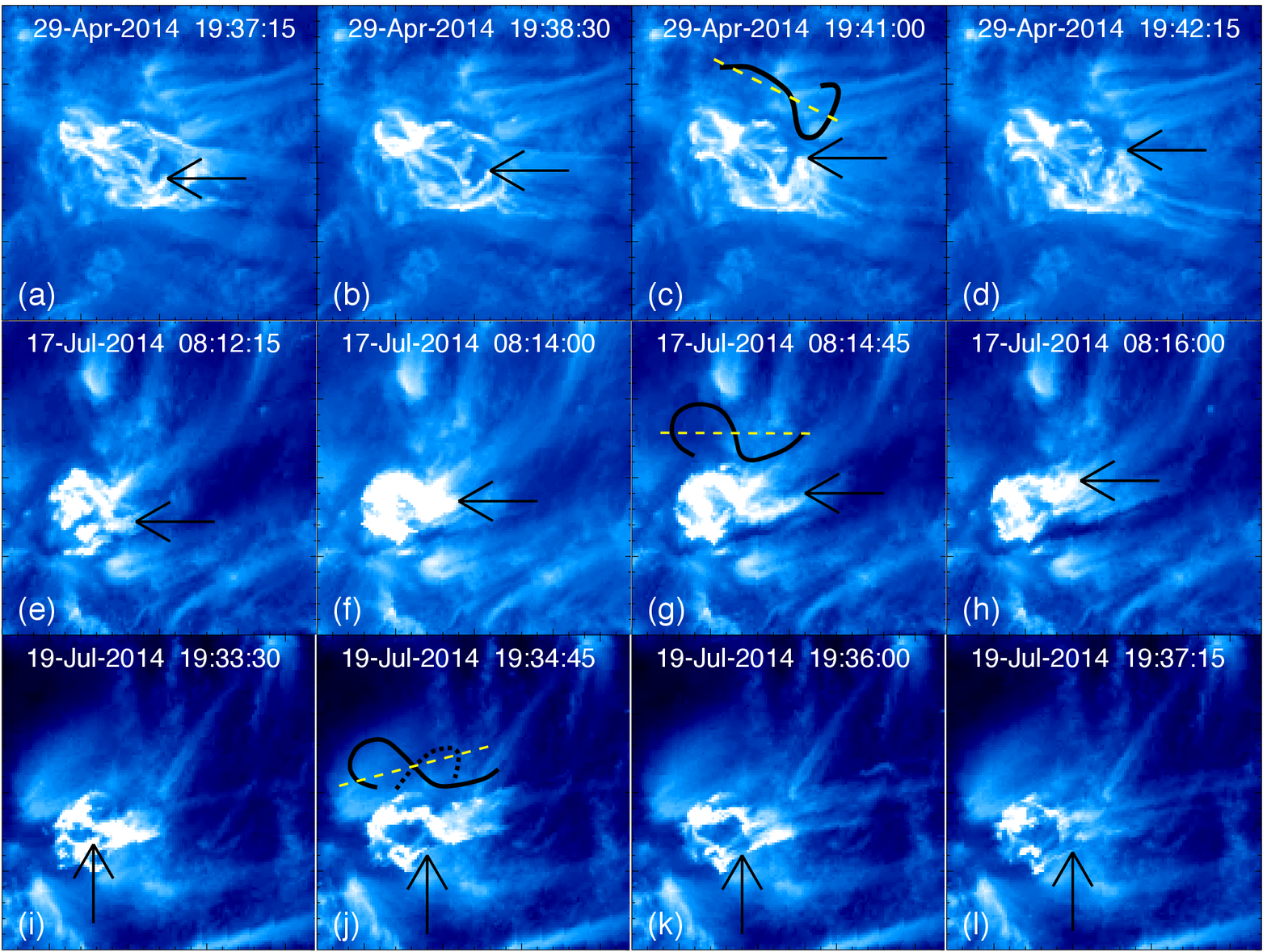}
\caption{STEREO-A EUV (171\,{\AA}) images of the jet evolution in event 1 (upper row), event 2 (middle row) and event 3 (lower row). The arrows mark the traced rotating spire. The schematic drawings resemble the observed helical configuration. (\href{https://figshare.com/s/60d9ba12ef9f5a2306cd}{An animation of this figure is available.})  \label{fig:f6}}
\end{figure*}

Figure \ref{fig:f6} shows the evolution of the EUV jets in high-cadence 171\,{\AA} images. The figure indicates that jets in all three examined events exhibit twisted structure (see also animation). A close examination of the 171\,{\AA} movies reveals counterclockwise rotation (when viewed from above) in the erupting jets. In a five-minute period 19:37--19:42\,UT on April 29 (Figure \ref{fig:f6}(a--d)), the inner spire (marked by a horizontal arrow) shows $\lesssim$1/4 of full (2$\pi$) rotation - the top end of the spire (a tracking feature) moves northward. This translates to the angular speed of $\sim$0.3$^{\circ}$\,s$^{-1}$. For a comparison, \citet{she11} reported the angular speed of 0.6$^{\circ}$\,s$^{-1}$ in an unwinding jet in the north polar coronal hole by examining 304\,{\AA} images. The spire in the event on April 29 is embedded within an erupting loop which looks broken at 19:38\,UT (see Figure \ref{fig:f6}b). A schematic drawing in Figure \ref{fig:f6}c resembles the observed helical-like configuration of the inner spire (solid line) rotating around its axial axis (dashed line). Four minutes later, at 19:46\,UT, the same spire becomes again bright and shows another rotation ($\lesssim$1/4 of a full turn). At 19:46\,UT the spire is viewed edge-on and at 19:48\,UT it is face-on (see animation - not shown in Figure \ref{fig:f6}). Now, the spire rotates at a faster rate ($\sim$0.75$^{\circ}$\,s$^{-1}$). The animation further shows that the spire has a shape of inverted omega and appears to be in a close contact with a wave-like open field line. In a four-minute period 08:12--08:16\,UT on July 17 (Figure \ref{fig:f6}(e--h)), the spire shows $\sim$1/4 of a full rotation with a tracking feature (top end of the spire) moving northward. A drawing in Figure \ref{fig:f6}g resembles the observed helical-like configuration of the spire. The approximate angular speed of the rotating plasma in the jet is $\sim$0.4$^{\circ}$\,s$^{-1}$. Note, the 5-min 195\,{\AA} difference images indicate the spire rotation in the July 18 15:20\,UT jet associated with the fourth ion injection in event 2. Figure \ref{fig:f6}(i--l) shows rotation of a spire (marked by a vertical arrow) during a four-minute period 19:33--19:37\,UT on July 19. A tracking feature moves southwest. A drawing in Figure \ref{fig:f6}j outlines the observed helical-like configuration with two crossing threads where a dotted curved line represents a backside thread. The animation shows a rotation of the same spire in another four minutes (19:39--19:43\,UT) when the spire dimmed. An inverted-Y shape of the dimmed spire is clearly seen. Here a tracking feature moves northward. Thus, within ten minutes the spire rotated in a half turn which indicates the angular speed of $\sim$0.3$^{\circ}$\,s$^{-1}$.

The jet on July 17 \added{08:09\,UT} shows an obvious helical motion only within $\sim$5\,minutes after the flare; in the event on April 29 and July 19 the helical shape along the jets is more persistent ($\sim$10\,minutes). Thus, the previous observations of $^3$He-rich solar sources with lower temporal resolution (above $\sim$5 or 10\,minutes) might have not seen a twisted topology in the jets. \citet{moo13} have reported blowout helical jets in about half (50/109) of all observed X-ray jets in polar coronal holes. Similar results have been presented in a survey of EUV jets \citep{nis09} where 31 out of 79 polar coronal-hole jets showed a helical configuration. The authors pointed out that the helical structure might not be resolved if the jet is very narrow or if the helical phase is too short. However, there is evidence that small-scale rotations may be present also in standard jets \citep{moo13,par15}. Statistical investigations of low-latitude coronal jets have not been performed owing to difficulties in examining the jet morphology against bright coronal structures \citep{nis09}.

\begin{figure*}
%\figurenum{7}
%\epsscale{1.}
\plotone{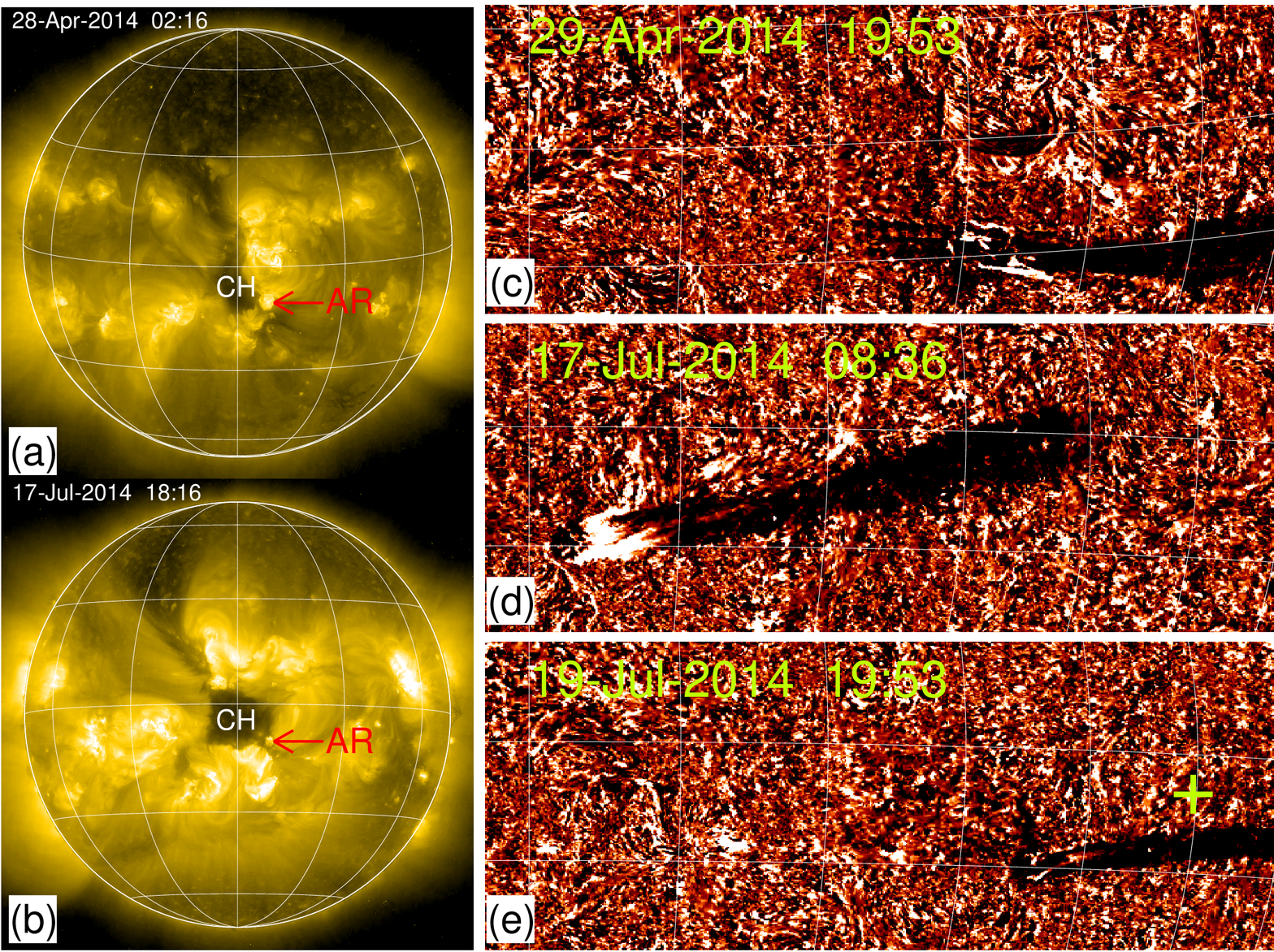}
\caption{(a--b) STEREO-A 284\,{\AA} EUV images of the solar disk. The source active regions (ARs) on the border of a coronal hole (CH), in event 1 (panel a) and events 2 and 3 (panel b), are marked by red arrows.\deleted{The EUV images are shown at times when the coronal hole was near the central meridian as viewed from STEREO-A.} The heliographic longitude-latitude grid has 30$^{\circ}$ spacing. (c--e) STEREO-A 304\,{\AA} EUV 10-min running difference images of the area on the solar western hemisphere for event 1 (panel c), event 2 (panel d) and event 3 (panel e). White (black) indicates increasing (decreasing) emission over the difference time interval. Green pluses mark the STEREO-A magnetic foot-point. The heliographic longitude-latitude grid has 10$^{\circ}$ spacing. \label{fig:f7}}
\end{figure*}

The source active region in event 1 emerged at the border of a near-equatorial coronal hole on April 26 $\sim$05\,UT as seen in EUV 304\,{\AA} images. Note that photospheric magnetic field observations, where emerging active regions are usually well observed, are not available from the backside of the Sun. The source region in events 2 and 3 is also located on the equatorial coronal hole border and emerged on July 16 around 12\,UT. The coronal holes, the regions with magnetic field open to the heliosphere, were most clearly seen in 284\,{\AA} EUV images \replaced{(see Figures \ref{fig:f7}a and \ref{fig:f7}b)}{shown in Figure \ref{fig:f7}(a--b)}. \added{The EUV images are shown at times when the coronal hole was near the central meridian as viewed from STEREO-A.} Figure \replaced{\ref{fig:f7}a and \ref{fig:f7}b}{\ref{fig:f7}(a--b)} demonstrates that the solar sources were relatively (compared to other bright areas on the disk) small and compact regions. Figure \ref{fig:f7}(c--e) shows STEREO-A 304\,{\AA} running difference images $\sim$30-min after the flare onset time in all three investigated events. We can see that associated jets (dark stretches) show significant non-radial expansion aligned with the STEREO-A foot-point longitude. The jets were highly collimated and remarkably long; the projected length is $\sim$40--60$^{\circ}$ longitude (or $\sim$490--735\,Mm). This indicates how coronal field lines were broadly extended from the jet locus. The wide spreading of open coronal field lines (up to 60$^{\circ}$) in $^3$He-rich SEP sources has been previously noted from the field model extrapolations \citep{wie13}. Recall from Table \ref{tab:tab2} that the STEREO-A foot-point was separated from the source region in event 2 by $\sim$42--65$^{\circ}$. The STEREO-B, where event 2 was also seen more weakly, had the foot-point placed even farther (86$^{\circ}$) from the source region. 

\section{Discussion} \label{sec:dis}

We have examined three $^3$He-rich SEP events that showed high enrichment in both $^3$He and heavier ions. The two events exhibited extremely high ion fluences. The solar sources of the examined events were associated with giant blowout jets that were very prominent in cooler chromospheric 304\,{\AA} EUV images. The jets originated at the boundary of a near-equatorial coronal-holes. It is interesting that all these events contained a mini filament at their solar source. A ring-shape filament has been recently reported in a solar source for the intense 2014 May 16 $^3$He-rich SEP event \citep{mas16} with very large $^3$He \added{and Fe} abundance\added{s} ($^3$He/$^4$He$\sim$\replaced{15}{14.9} at 0.5--2\,MeV\,nucleon$^{-1}$\added{, Fe/O$\sim$2.1 at 0.32--0.45\,MeV\,nucleon$^{-1}$}) \citep{nit15}. Note, the Fe fluences (at 0.1--1\,MeV\,nucleon$^{-1}$) in the May 16 event are more than an order of magnitude lower than in event 2; the $^3$He fluences are equally intense as in event 2. A filament in a solar source of $^3$He-rich SEPs has been reported in the earlier study by \citet{kah87} by inspecting H$\alpha$ film patrols. The authors examined 12 events and found a circular filament in two sources with a quite high $^3$He/$^4$He ratio ($\sim$2.0--2.4 at 1\,MeV\,nucleon$^{-1}$) \citep{rea85}. A twisted flux rope at the jet source, which may be carried by a mini-filament, has been considered as a triggering mechanism for blowout helical jets \citep{moo10,moo13,mor13,arc13}. The twist stored in the base arch, that has lots of confined free energy, is transformed into the open field during magnetic reconnection.  

The most intriguing feature in the $^3$He-rich SEP events under study is a helical EUV jet seen in the corresponding solar sources. Several numerical works have recently shown that the untwisting motion of the helical blowout jet produces large amplitude Alfv\'en waves propagating upward along newly reconnected open field lines \citep{par09,tor09,lee15}. These waves could be traced by a helical structure in jets \citep{moo15} as the plasma is forced to move with the magnetic field. \deleted{The observational evidence of Alfv\'en wave generation has been reported a decade ago by \citet{cir07} analyzing large X-ray jets in polar coronal holes.}The propagation of Alfv\'en waves has been revealed by examining space-time diagrams across the near equatorial active region blowout jets in chromospheric Ca II He ($<$0.02\,MK) \citep{nis08,liu11} and 171\,{\AA} \citep{sch13} lines.

\replaced{\citet{rog03} have suggested that the helical flow of magnetized plasma may be an efficient amplifier of Alfv\'en waves in the solar atmosphere. This may perhaps lead to the enhanced}{A} gyro-resonant wave-particle interaction (referred as stochastic acceleration) \added{has} often \added{been} employed in acceleration models for $^3$He-rich SEPs. Specifically, the models of cascading Alfv\'en waves may account for heavy and ultra-heavy ion acceleration \citep{mil98,zha04,eic14,kum17}. The models assume that Alfv\'en waves are generated during the magnetic reconnection by relaxation of twisted non-potential magnetic fields \citep{mil98}. \added{Cascading toward shorter length scales the waves resonate with ions of increasing gyro-frequency.} These models appear to approximate well the low-energy heavy ion spectra \citep{mas02}. The models of $^3$He acceleration involve plasma waves generated around the $^3$He cyclotron frequency \citep{fis78,tem92}. These waves are assumed to be generated by an electron current or energetic electron beams. \citet{liu06} included in their model the resonance with plasma wave turbulence in the proton cyclotron branch, which reasonably well reproduces observed $^3$He and $^4$He spectra. The authors discussed the generation of these waves via coupling with low-frequency Alfv\'en waves. \citet{fle13} have suggested that the rate of stochastic particle acceleration is greatly enhanced when the turbulence is helical (i.e. when the magnetic field lines are twisted). The authors further suggested that a helicity-driven electric field can be responsible for anomalous fractionation of $^3$He and heavy elements, forming a seed population which eventually undergoes standard stochastic acceleration. The \citet{gor14} model shows that the electric field in twisted coronal loops can effectively (with reduced losses) accelerate electrons and protons. 

Not all reported $^3$He-rich SEP events have been found to be associated with coronal jets, though it is generally accepted that it is mainly due to observational constraints (like resolution, observing angle). Using {\sl Solar and Heliospheric Observatory} 195\,{\AA} images \citet{wan06} found EUV jet in 13 of the 22 cases. The authors argued that more jets would be identified with higher temporal resolution ($<$12\,minutes) observations. Remaining events in their survey include ejections/surges (5 cases) or an amorphous flare brightening (4 cases). \citet{nit06} have used higher cadence {\sl Transition Region and Coronal Explorer} 171, 195\,{\AA} data and identified EUV jets in a slightly higher fraction of events (17 out of 25). Remaining events in their study involve loop brightening (5 events) or filament eruptions (3 events). Using even higher-cadence multi-wavelength {\sl Solar Dynamics Observatory} EUV images, \citet{nit15} reported EUV jets only in about half of the cases (13 out of 29). The authors pointed out that when a primary activity is a wider eruption (12 cases) or an EUV wave (4 cases), the jet if present, may be overlooked. In their study, at least one event with the EUV wave starts as a jet. Using the STEREO 195\,{\AA} data, \citet{buc16} reported a somehow weaker association with EUV jets. The authors identified jets at least in 10 of the 32 cases; other activity includes small brightening and EUV waves. It appears that an identification of a type of activity in the solar source may be biased by the selection of the events. For instance, \citet{nit15} have included in their study only relatively intense $^3$He-rich SEP events, while \citet{buc16} have examined also weak events with the ion intensities near the detection threshold.

From 11 jets displayed in \citet{wan06},\deleted{two show clear twisted configuration,} \replaced{six}{eight} are probably blowout (\added{two} with \deleted{no}clear twist\added{ed configuration}) as they show broad and structured spire, two are ambiguous, and only one is a standard jet with a point-like brightening at the base and well-collimated spire. In addition, one more EUV event classified as an ejection (2000 January 6) is probably a blowout jet that exhibits the highest $^3$He enrichment ($^3$He/$^4$He $\sim$33.4 at 0.386\,MeV\,nucleon$^{-1}$) ever reported. \citet{nit06} display two EUV jets that are a standard type, and \citet{nit15} show examples of two EUV jets in a detail view that are the blowout. A statistical study of \citet{buc16} shows images of EUV waves well after a jet time. Several case-studies display jets in $^3$He-rich SEP sources \citep{nit08,buc14,buc15,che15}. The only example of an X-ray jet associated with $^3$He-rich SEPs is probably also a blowout type \citep{nit08}. One recurrent jet in a study of \citet{buc14} appears to be a standard jet, but another is probably blowout type. The jets in two $^3$He-rich SEP events, that trigger EUV wave, were blowout type (one with a helical motion) as could be inferred from the animations in \citet{buc15}. The jet in \citet{che15} was also blowout with a curtain-like spire. Finally, a helical \replaced{EUV}{blowout} jet was clearly seen in the solar source of above mentioned 2014 May 16 event \citep{inn16}. \deleted{Thus, the previous studies indicate that coronal jets associated with $^3$He-rich SEPs were predominantly blowout type.}

\replaced{In summary, the observations of helical jets in the solar source of the investigated $^3$He-rich SEP events appear to directly support models of wave-particle interactions. We have examined cases with a filament and a helical structure well seen in the particle source. Further studies with high-resolution observations are needed to reveal configuration in the sources of ordinary $^3$He-rich SEP events.}{Although many jets in previous studies on $^3$He-rich SEPs appear to be a blowout type, further high spatio-temporal resolution observations are needed to establish the association and reveal a helical shape in source flares. The helical jets in the examined $^3$He-rich SEP sources may be a distinct feature that produces events greatly enriched both in $^3$He and Fe. In view of the present models, Alfv\'en waves of the blowout jets may account for the Fe enrichment. The $^3$He enhancement may be related to the twisted configuration of the source flare, for instance via enhanced production of the energetic electron beams.}

\acknowledgments
This work was supported by the Deutsche Forschungsgemeinschaft (DFG) under grant BU 3115/2-1 and Max-Planck-Gesellschaft zur F\"orderung der Wissenschaften. The STEREO SIT and SECCHI are supported by the Bundesministerium f\"ur Wirtschaft through the Deutsches Zentrum f\"ur Luft- und Raumfahrt (DLR) under grant 50 OC 1301. Work at Johns Hopkins Univ. APL was supported by NASA grant NNX17AC05G/125225, and Univ. of California Berkeley grant 00008934. Work at JPL was supported by NASA. RGH acknowledges the financial support of the University of Alcal\'a under project CCG2015/EXP-055 and the Spanish MINECO under project ESP2015-68266-R. Nariaki Nitta's work was supported by NFS grant AGS-1259549. We thank Linghua Wang for providing calibrated STE/STEREO electron data.

\end{document}